\documentclass[11pt,twoside]{article}

\usepackage{asp2006}
\usepackage{epsf}
\usepackage{lscape}
\usepackage{amsbsy}
\usepackage{amssymb}

\begin{document}
\title{The Prototype of the\\Small Synoptic Second Solar Spectrum Telescope (S$^5$T)}

\author{Frans Snik,$^1$ Radek Melich,$^{2,3}$ and Christoph Keller$^1$}

\affil{$^1$Sterrekundig Instituut Utrecht, Princetonplein 5, 3584 CC Utrecht, the Netherlands}
\affil{$^2$Institute of Plasma Physics, Academy of Sciences of the Czech Republic}
\affil{$^3$Department of Optical Diagnostics, Skalova 89, 51101 Turnov, Czech Republic}


\begin{abstract}
We present the design and the prototype of the Small Synoptic Second Solar Spectrum Telescope (S$^5$T), which can autonomously measure scattering polarization signals on a daily basis with large sensitivity and accuracy. Its data will be used to investigate the nature of weak, turbulent magnetic fields through the Hanle effect in many lines. Also the relation between those fields and the global solar dynamo can be revealed by spanning the observations over a significant fraction of a solar cycle. The compact instrument concept is enabled by a radial polarization converter that allows for ``one-shot'' polarimetry over the entire limb of the Sun. A polarimetric sensitivity of $\sim$$10^{-5}$ is achieved by minimizing the instrumental polarization and by FLC modulation in combination with a fast line-scan camera in the fiber-fed spectrograph. The first prototype results successfully show the feasibility of the concept.
\end{abstract}


\section{Science Case}
The observation of weak, turbulent magnetic fields on the Sun is challenging for many reasons. First of all, the Zeeman effect is virtually blind to these fields and therefore they do not show up in conventional magnetograms. The discovery of the Second Solar Spectrum \citep{sss} paved the road for a complementary magnetic field diagnostic through the Hanle effect. The analysis by \citet{TBnature} based on various observations of the Second Solar Spectrum and extensive modeling of the action of the Hanle effect revealed the presence of ubiquitous weak magnetic fields ($\sim$100 G), entangled at sub-granular length scales. The apparent absence of variation of the scattering polarization signals with the solar cycle could be evidence for a local origin of these turbulent magnetic fields through granular dynamo action. Also, MHD models by \citet{voeglerdynamo} conjecture the existence of such a local dynamo in the solar photosphere. However, observations by \citet{Biandavar} clearly show significant variations in several lines of the Second Solar Spectrum with the solar cycle. Also, the polarimetric calibration accuracy of the currently available data-sets of the Second Solar Spectrum is insufficient for carefully comparing results obtained on different days or with different instruments and to unambiguously investigate the relation of the weak, turbulent fields to the global solar dynamo. Moreover, more extensive statistics of scattering polarization signals is required in order to develop a more complete picture of solar magnetism.\\
Therefore the need for a dedicated instrument to measure the Second Solar Spectrum accurately is obvious. It is required that such an instrument can reach a polarimetric sensitivity of $\sim$$10^{-5}$ in order to distinguish the weak signals of the Second Solar Spectrum and that it is able to reproduce the polarimetric scale at an accuracy level of $\sim$2\%. The instrument needs to be autonomous and fully automated in order to furnish daily measurements without interfering with operations at a major solar facility. We choose to operate in the wavelength range 415-465 nm, where a large concentration of prominent lines and features in the Second Solar Spectrum is to be found \citep{atlas}.

\section{The Theta Cell}
Traditionally, the Second Solar Spectrum is obtained by positioning a straight spectrograph slit at $\mu=0.1$ in a quiet region of a highly magnified solar image. A large-aperture telescope is necessary in order to obtain a sufficient photon flux for precision polarimetry, as it is impossible to average over large fractions of the solar limb within one observation. Such averaging leads to cancelation of the scattering polarization signals, since those are oriented parallel to the local limb direction, i.e. azimuthally on the solar disk.\\
Utilizing a passive twisted nematic liquid crystal component known as a radial polarization converter or ``theta cell'' \citep{thetacell}\footnote{See also \texttt{www.arcoptix.com/radial\_polarization\_converter.htm}.}, it is possible to convert the polar coordinate system of scattering polarization on the solar disk to one global Cartesian system. In other words, the local direction of Stokes $Q$ parallel to the limb is transformed everywhere into an overall direction of Stokes $Q$, which in turn can be measured by a single linear polarimeter.\footnote{The theta cell is composed of liquid crystals that rotate the direction of linear polarization by waveguiding. This effect is achromatic for wavelengths larger than the typical waveguide dimensions of the liquid crystals, in contrast to liquid crystal half-wave retarders, which are inherently chromatic. We verified experimentally that the theta cell rotates linear polarization from its circular to its linear geometry within $\pm$1$^\circ$ for wavelengths 400-750 nm.} Note that with this configuration only one linear Stokes parameter needs to be determined, instead of both Stokes $Q$ and $U$ as is the general case for the azimuthal orientation of linear scattering polarization signals.
With such a theta cell in prime focus, only a small-aperture telescope is necessary to deliver the photon flux for measuring the Second Solar Spectrum averaged over the entire limb. Also the need for large magnification and hence a long focal length is no longer there to make to limb curvature much larger than the slit length. After selection of the limb regions of the synoptic solar image, all the relevant light can be polarimetrically analyzed and consecutively fed into a spectrograph. Therefore, the use of a theta cell reduces the required telescope size dramatically and gives rise to the design of the Small Synoptic Second Solar Spectrum Telescope (S$^5$T).\footnote{Also observations of various other astronomical targets with centrosymmetric single scattering geometries can be simplified by the use of a theta cell: e.g. the K corona, reflection nebulae, circumstellar disks and exoplanets.}

\section{Optical Design}
\begin{figure}[!h]
\begin{center}
\plotone{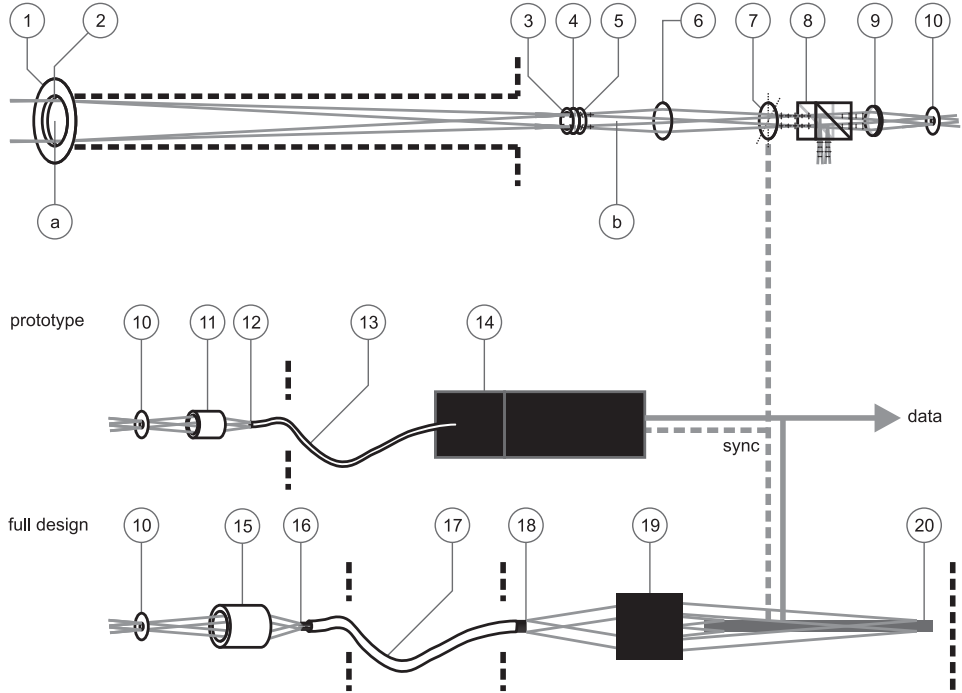}
\caption{Conceptual design of the S$^5$T, both for the prototype and the full instrument. The individual components are discussed in the text.}
\label{optdes}
\end{center}
\end{figure}

The optical design of the S$^5$T is schematically depicted in figure \ref{optdes} for both the realized prototype and for the full instrument. The design is optimized for precision polarimetry with a minimal amount of instrumental polarization.

\vspace{-15 pt}

The sequential optical components are:\vspace{5 pt}
\begin{list}{\arabic{enumi}.}{\usecounter{enumi}\setlength{\topsep}{0pt}\setlength{\parsep}{0pt}\setlength{\itemsep}{0pt}}
\item A reflective ring reducing the thermal stresses in the mounting of the objective lens.
\item A 5 cm fused silica singlet objective lens with a focal length of 1 m. This configuration has minimal instrumental polarization because of the circular symmetry and the low amount of stress birefringence of fused silica. The F/20 beam keeps the chromaticity in prime focus tolerable for course field selection and its \'etendue allows for considerable demagnification to fiber-feed the spectrograph as described below.
\item The heat stop close to prime focus reflects and absorbs most of the unwanted sunlight in the center of the disk. Only course field selection is performed here, due to the chromaticity of the first image plane.
\item A dichroic color filter rejects unwanted UV and red-IR light in order to reduce stray-light issues and protect the liquid crystal devices from harmful radiation.
\item The theta cell is located in prime focus and transforms the azimuthally oriented scattering polarization signals on the $\sim$1 cm large solar disk image into one global direction of linear polarization.
\item A second fused silica lens collimates the beam and images the pupil onto the polarimeter (components 7 and 8), in order to ensure identical polarization modulation for all field points.
\item A half-wave ferro-electric liquid crystal (FLC) modulates the $+Q$ and $-Q$ polarization directions at kHz rates.
\item A polarizing beam-splitter transmits only one polarization direction and transforms the polarization modulation of the FLC into an intensity modulation.
\item A third lens creates a telecentric second image plane. Since the polarization analysis has been performed, this lens can be polarimetrically unfavorable. A custom ``antichromatic'' doublet is designed to create an achromatic second image, compensating for the chromatic effects of the first two lenses. In the prototype a regular achromat is used.
\item A precise field selector is located in the second image plane. It transmits only the outer $\sim$10\arcsec~of the solar disk. Only the positions within $\pm$45$\deg$ from the solar poles are selected, as the other areas are frequently visited by active regions. From the field selection there are two options for feeding the light into a spectrograph: feeding the pupil into a single fiber (as is done for the prototype) or reimaging the selected field onto a circular fiber bundle (which is the preferred option for the final instrument).\vspace{10 pt}
\item For the prototype, a microscope objective creates a collimated beam again.
\item The pupil image of $\sim$1 mm is fed into a multimode fiber with a 400 $\mu$m core diameter. This reduces the effective telescope aperture to $\sim$2 cm.
\item The fiber transfers a part of the pupil into the spectrograph.
\item The spectrograph for the prototype is a HR4000 model from Ocean Optics. The spectral range is 400-485 nm and a 5 $\mu$m slit ensures a spectral resolution of $\sim$0.5\AA. Its 4k line-scan camera is synchronized with the FLC.\vspace{10 pt}
\item Because so much light gets lost on feeding the fiber in a pupil image and feeding the spectrograph with a circular fiber core through a narrow slit, a different concept is envisaged for the final instrument. A fourth lens reimages the selected field.
\item The narrow light ring is sampled by a circular fiber bundle containing multimode fibers with a core size of 50 $\mu$m.
\item The fiber bundle transports the light to the spectrograph elsewhere.
\item The fibers are reformatted to form a quasi-slit of $\sim$1 mm high. A telecentric microlens array matches the F/\# of the fibers to the F/\# of the spectrograph. A slit is positioned at the back end of the microlens array in order to increase the spectral resolution.
\item A demagnifying Czerny-Turner spectrograph.
\item In the end, the spectrum will be recorded by a commercially available 8192$\times$96 pixel line-scan camera. The fast camera will be synchronized to the FLC modulation. The camera is able to fully bin the signals in the spatial direction, thereby fully compressing the light from the entire solar limb to one final signal. With a spectral range of 415-465 nm, a spectral resolution of $\sim$0.15 \AA~can be attained.
\end{list}
\vspace{10 pt}
No calibration components have been developed for the prototype. In order to ensure the accuracy and stability of the final instrument, a number of calibration procedures are foreseen: A rotating polarizer can be inserted at position b.~of figure \ref{optdes} in order to calibrate the polarimeter. In order to measure the instrumental polarization due to the two fused silica lenses and to check for degradation of the theta cell, also a rotatable polarizer can be inserted before the objective lens at position a. By using the spectrograph in imaging mode, i.e. taking spectra of each individual fiber in the bundle, the polarization rotation properties of the theta cell can be measured.

\section{Prototype Results}
\begin{figure}[!h]
\begin{center}
\plotone{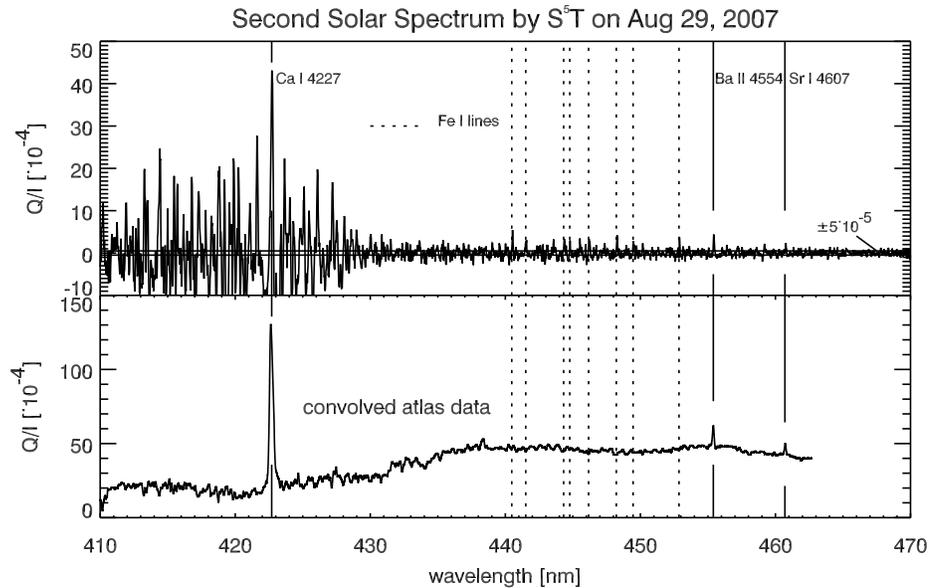}
\caption{First data from the S$^5$T prototype on Aug. 29, 2007 at Sonnenborgh observatory in the center of Utrecht. The total exposure time was 4 minutes at 250 Hz modulation rate. The noise level between 456 and 470 nm has an RMS of $6\cdot10^{-5}$. The sensitivity towards the far blue was limited by the low photon counts.}
\label{S5Tresults}
\end{center}
\end{figure}

The prototype of S$^5$T was designed and constructed within the first half of 2007. The optical components 3-12 of figure \ref{optdes} were aligned within a 60 cm long box with an attached tube of 1 m length to hold the objective lens. The entire assembly was attached to a tripod with a motorized equatorial mount. Pointing was performed manually, using a quad cell behind a separate pinhole camera as a reference signal. The prototype was set-up at the old Utrecht observatory ``Sonnenborgh'', next to the heliostat that has been used for pioneering work on the (``first'') solar spectrum \citep{Minnaert}.\\
The first results are presented in figure \ref{S5Tresults} and compared to the Second Solar Spectrum from the atlas of \citet{atlas} after smearing with a 0.5 \AA~Gaussian. The data has been dark-subtracted and Fourier-filtered in order to remove polarized spectral fringes originating from the liquid crystal devices. It is clear that within 4 minutes the prototype reached a polarimetric sensitivity at the $10^{-5}$ level and observed the scattering polarization signals of the Ca~{\sc{i}}~4227, the Ba~{\sc{ii}}~4554 and the Sr~{\sc{i}}~4607 lines. Also a number of deep Fe~{\sc{i}} lines show up in the polarimetric data, whereas they are absent in the atlas. Initial tests on the data show that these signals can not be due to bias drift or coupling of the fringe pattern to the camera non-linearity cf. \citet{nonlin}. In the future, also ``null'' measurements have to be taken by rotating the FLC to modulate Stokes $U$ only. This way, the Fe~{\sc{i}} lines' signals can be determined to be instrumental or solar in origin and also the spectral fringes can be measured independently.

\section{Outlook}
The final instrument is fully funded by NOVA (The Dutch research school for astronomy) as of November 2007 and will be designed, constructed, tested and commissioned in the course of the next two years. It will be mounted at the fourth instrument platform of SOLIS at Kitt Peak, AZ, USA. In the mean-time, the prototype will be used for further tests. It will be updated in order to be able to observe the polarization of the flash spectrum during the 2009 solar eclipse over China.

\newpage

\acknowledgements
The prototype of S$^5$T was funded by SOZOU. The engineering and manufacturing has been done by the Instrumentele Groep Fysica of Utrecht University. This research project has been supported by a Marie Curie Early Stage Research Training Fellowship of the European Community's Sixth Framework Programme under contract number MEST-CT-2005-020395. FS acknowledges travel funds to the SPW5 conference from LKBF.


\end{document}